\documentclass[%
 reprint,
 superscriptaddress,
 showpacs,preprintnumbers,
 amsmath,amssymb,
 aps,
 prb,
 longbibliography,
lengthcheck,%
]{revtex4-1}

\usepackage{graphicx}
\usepackage{dcolumn}
\usepackage{bm}
\usepackage{color}
\usepackage{ulem}

\begin{document}

\preprint{APS/123-QED}

\title{
``Switching'' of Magnetic Anisotropy in a fcc Antiferromagnet\\ with Direction-Dependent Interactions
}

\author{Hiroaki Ishizuka}
\affiliation{
Kavli Institute for Theoretical Physics, University of California, Santa Barbara, California 93106, USA
}

\author{Leon Balents}
\affiliation{
Kavli Institute for Theoretical Physics, University of California, Santa Barbara, California 93106, USA
}

\date{\today}

\begin{abstract}
Direction-dependent anisotropic exchange is a common feature of magnetic systems with strong spin-orbit coupling.  Here we study the effect of such exchange upon macroscopic magnetic anisotropy for a face-centered-cubic model.   By several theoretical techniques, we show that, both in the paramagnetic and ordered phases, the magnetic anisotropy is induced by fluctuations.   Moreover, the magnetic anisotropy differs in the paramagnetic and ordered phases: in the paramagnetic phase the susceptibility is maximum along the $\langle111\rangle$ directions, while the magnetic moments orient along $\langle110\rangle$ or $\langle100\rangle$ in the ordered phase.  We suggest that such ``anisotropy switching'' can be a common feature of strongly spin-orbit coupled magnets.  
\end{abstract}


\maketitle

Insulating magnetic compounds with heavy elements, such as $4d$ and $5d$ transition metal ions, are of considerable interest in the study of quantum magnetism. In these materials, strong electron correlation and spin-orbit coupling often lift the orbital degrees of freedom, realizing ``pure'' spin systems with effective pseudo-spins comprised of by entangled microscopic spins and orbitals.  Generically, such strong spin-orbit coupling gives rise to direction-dependent exchange interactions.  The presence of such interactions potentially affects the magnetic properties in a significant way, possibly stabilizing exotic phases, such as spin-liquid states.~\cite{Kitaev2006,Jackeli2009} Experimentally, several iridates~\cite{Cao2003,Singh2010,Liu2011,Choi2012,OMalley2008,Singh2012,Modic2014,Takayama2014,Biffin2014} and double perovskites~\cite{Stitzer2002,Wiebe2002,Wiebe2003,Yamaura2006,Cussen2006,Erickson2007,deVries2010,Aharen2010} have been studied in these contexts.

In this paper, we theoretically study a model of an antiferromagnet on a face-centered cubic (fcc) lattice with direction-dependent interactions.  This model may apply directly to cubic double perovskites, but also serves as a representative example of the general class of frustrated systems with directional couplings.   We observe that, at low temperature, the primary role of the directional interactions is to partially lift the ground state degeneracy of the isotropic problem, selecting a particular subset of ground states. In addition to this ``direct'' lifting of the ground state manifold, quantum fluctuations also affect the ground state selection -- a phenomena known as ``order by disorder''.~\cite{Villain1980,Shender1982}

At the same time, the directional couplings and fluctuations also affect the magnetism in the high temperature paramagnetic phase, and in particular the magnetic susceptibility and anisotropy.   Importantly, the nature of fluctuations in the high temperature phase is different from those in the ordered phase at low temperature.  In the high temperature phase, since the spins are strongly disordered, correlations between nearby spins are dominant. On the contrary, long wavelength fluctuations tend to be more important in ordered phases. Hence, the contribution of fluctuations in these two regimes may result in qualitatively different behavior.

We show that this is indeed the case in fcc antiferromagnets.  By a combination of the Luttinger-Tisza method, spin wave theory, and the high-temperature expansion, we show that the magnetic anisotropy ``switches'' between the paramagnetic and the ordered phases.  Specifically, at low temperature in the ordered phase, either the $\langle100\rangle$ or $\langle110\rangle$ directions are favored, depending on the sign of the directional exchange coupling, while in the paramagnetic phase, the $\langle111\rangle$ direction is favored.  We show that quantum fluctuations dominate in selecting the ground state, while thermal fluctuations drive the magnetic anisotropy in the paramagnetic phase.

The model we consider consists of spins on the fcc lattice with direction-dependent nearest-neighbor (NN) exchange and isotropic second-neighbor Heisenberg interaction,
\begin{eqnarray}
H &=& H_0+H_h \label{eq:H}
\end{eqnarray}
with
\begin{eqnarray}
H_0 &=& J_1 \sum_{\langle i,j\rangle} {\bf S}_i\cdot {\bf S}_j - J_2 \sum_{\left[ i,j\right]} {\bf S}_i\cdot {\bf S}_j\nonumber\\
  && \quad\quad\quad+ 2J_3 \sum_{\langle i,j\rangle} ({\bf S}_i\cdot{\boldsymbol \delta}_{ij})({\bf S}_j\cdot{\boldsymbol \delta}_{ij})\label{eq:H0}\\
H_h &=&  -{\bf h}\cdot\sum_i {\bf S}_i.
\end{eqnarray}
Here, ${\bf S}_i=(S_i^x,S_i^y,S_i^z)$ is the spin operator for the localized spin at $i$th site and ${\bm \delta}_{ij}$ is the vector connecting $i$th and $j$th sites. We choose the standard convention so that the conventional unit cell of the fcc cube has unit length. The first and the third sum in Eq.~(\ref{eq:H0}) is taken over all the NN sites and the second one is for second-neighbor sites. $J_1$ and $J_2$ are the NN and second-neighbor Heisenberg interaction, and $J_3$ is the direction dependent interaction. $H_h$ is the coupling of uniform magnetic field ${\bf h}=(h_x,h_y,h_z)$ to the spins. In the following, we focus on the case $J_1, J_2>0$.

\begin{figure*}
\includegraphics[width=\linewidth]{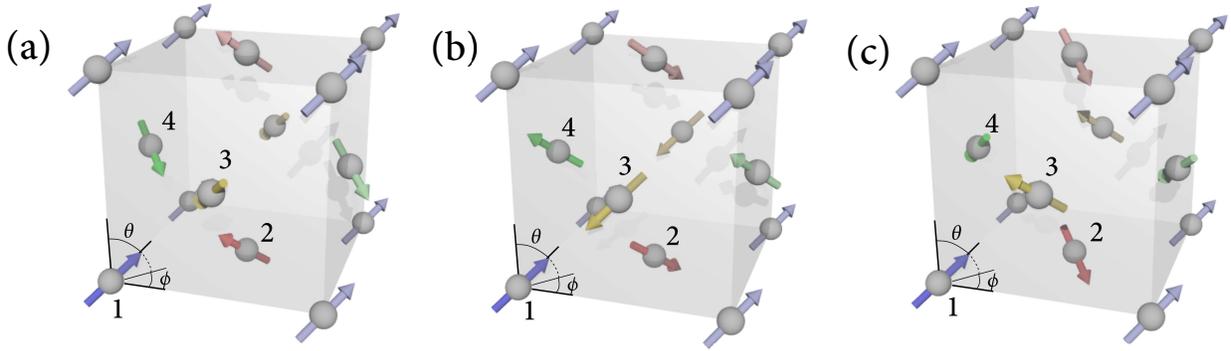}
\caption{
Schematic picture of the classical ground state for the model in Eq.~\ref{eq:H} with (a) $J_3>0$, and (b) double-$\bf q$ and (c) triple-$\bf q$ magnetic ground states for $J_3<0$. The dark colored spins indicate magnetic unit cell and the numbers are sublattice indices. The polar coordinate of the magnetic moments on $i$th sublattice, $\omega_i=(\theta_i,\phi_i)$ is given in Tab.~\ref{tab:fcc}.
} \label{fig:fcc}
\end{figure*}
\underline{Classical ground states:}
We first consider the magnetic ground states at ${\bf h}={\bf 0}$. When $J_3=0$, the classical ground state of the model in Eq.~(\ref{eq:H}) is given by a four sublattice configuration with the simple cubic magnetic unit cell and zero net magnetization. In Fourier space, these states are given by a linear combination of plane waves with wave numbers in the set ${\sf Q}= \{ {\bf q}_x,{\bf q}_y,{\bf q}_z \}$, with ${\bf q}_x =(2\pi,0,0)$, ${\bf q}_y=(0,2\pi,0)$, and ${\bf q}_z=(0,0,2\pi)$ (Ref.~\onlinecite{Henley1987}). To investigate the effect of $J_3$, we studied the classical ground state by using Luttinger-Tisza method.~\cite{Luttinger1946} In the Luttinger-Tisza method, the ground state is determined by the minimum eigenvalue of
\begin{eqnarray}
J({\bf q}) &=& \sum_{i} J_{i0} e^{-{\rm i}{\bf q}\cdot{\bf r}_{i0}},
\end{eqnarray}
where $J_{ij}$ is the $3\times3$ matrix that represents interaction between $i$th and $j$th spins. The spins are determined by a superposition 
\begin{equation}
\label{eq:1}
{\bf S}_i = \sum_{a=x,y,z} {\bf A}_a  e^{i {\bf q}_a\cdot {\bf r}_i},
\end{equation}
where $a$ is sum over ${\bf q}_a\in{\sf Q}$, and the vector coefficients ${\bf A}_a$ must be further constrained.  When $J_3=0$, $J({\bf q})$ is diagonal for arbitrary $\bf q$, with the minimum located at ${\bf q} \in {\sf Q}$.

\begin{table}
\begin{tabular}{c|ccc}
\hline
 & Fig.~\ref{fig:fcc}(a) & Fig.~\ref{fig:fcc}(b) & Fig.~\ref{fig:fcc}(c) \\
\hline
${\bf S}_1$ & $(\theta,\phi)$         & $(\theta,\phi)$         & $(\theta,\phi)$ \\
${\bf S}_2$ & $(\theta,\phi+\pi)$     & $(\pi-\theta,\phi)$     & $(\pi-\theta,-\phi)$ \\
${\bf S}_3$ & $(\pi-\theta,\pi-\phi)$ & $(\pi-\theta,\phi+\pi)$ & $(\theta,\phi+\pi)$ \\
${\bf S}_4$ & $(\pi-\theta,-\phi)$    & $(\theta,\phi+\pi)$     & $(\pi-\theta,\pi-\phi)$ \\
\hline
\end{tabular}
\caption{
Polar coordinate of the moments on different sublattice for the magnetic orders in Fig.~\ref{fig:fcc}(a)-(c).
}
\label{tab:fcc}
\end{table}

When $J_3>0$, the the coefficients must be chosen so that ${\bf A}_a$ is parallel to ${\bf q}_a$. The magnetic ground state is given by an arbitrary combination of these 3 modes that satisfies local constraint $|{\bf S}_i|=1$. An example is shown in Fig.~\ref{fig:fcc}(a). Defining polar coordinates $\theta$ and $\phi$ as shown in Fig.~\ref{fig:fcc}(a), arbitrary $\theta\in[0,\pi]$ and $\phi\in[-\pi,\pi]$ define the ground state manifold. It reduces to a two sublattice collinear structure for $\theta=0,\pi/2$ and $\phi=n\pi/2$ ($n=0,\cdots,3$), while in general it has four distinct sublattices for other sets of $(\theta,\phi)$. The direction of each spin ${\bf S}_i$ in polar coordinates is given in Tab.~\ref{tab:fcc}.

For small $J_3<0$, it is instead required that ${\bf A}_a$ is normal to ${\bf q}_a$. The magnetic ground state is given by an arbitrary combination of these 6 modes that satisfies local constraint $|{\bf S}_i|=1$. The solution consists of two manifolds of states as shown in Fig.~\ref{fig:fcc}(b) and \ref{fig:fcc}(c): a double-$\bf q$ state in which two out of the three components ${\bf A}_a$ are finite [Fig.~\ref{fig:fcc}(b)], and a triple-$\bf q$ state which all three components are finite [Fig.~\ref{fig:fcc}(c)].

The double-$\bf q$ state consists of two vector components, ${\bf q}_1$ and ${\bf q}_2$, which are elements of ${\sf Q}$. Figure~\ref{fig:fcc}(b) shows an example with ${\bf q}_1={\bf q}_z$ and ${\bf q}_2={\bf q}_x$. In this phase, the spin axis ${\bf A}_2$ is fixed parallel/antiparallel to ${\bf q}_1$, while ${\bf A}_1$ may be in an arbitrary direction.  The explicit directions of spins in the double-$\bf q$ state is given in Table~\ref{tab:fcc}, and an example indicated in Fig. ~\ref{fig:fcc}(b).  In the $(\theta,\phi)$ notation, the special values $\theta=0,\pi/2$ reduce to a single-$\bf q$ two-sublattice configuration, while other values with $\theta\in(0,\pi/2)$ gives a four-sublattice double $\bf q$ state.

Figure~\ref{fig:fcc}(c) shows an example of the triple-$\bf q$ state. In this state, the coefficients ${\bf A}_a$ are orthogonal to one another. The four sublattice spin pattern collapses to a double-$\bf q$ one for $\theta=0,\pi/2$, and a single-$\bf q$ state for $\phi=n\pi/2$ ($n=0,\cdots,3$). We note that with increasing $J_3<0$, the four-sublattice orders eventually become unstable.  The condition that the above solution is the energy minimum gives $J_2 > |J_3|/8$.   Hence, if $|J_3|$ is sufficiently small, the above-mentioned phases are expected to be stable.

\underline{Quantum order by disorder:} We next study how the quantum fluctuations lift the remaining degeneracy of the classical ground state manifold. We calculated the quantum correction to the classical ground state energy by spin wave analysis. Here, the unit of energy is taken as $J_1S$, where $S$ is the size of spin. The calculation were done numerically with $N_s=16^3$ magnetic supercells.

\begin{figure}
\includegraphics[width=\linewidth]{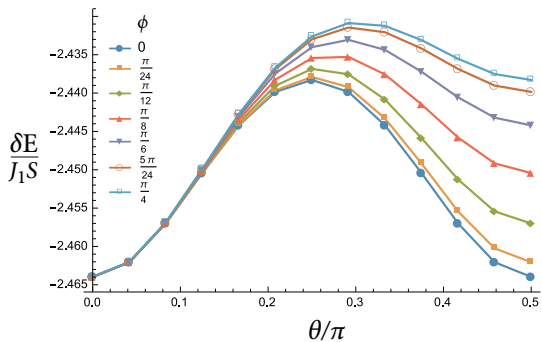}
\caption{
Quantum correction to the ground state energy for different ground states at $J_1=1$, $J_2=0.2$, $J_3=1$. $\theta$ and $\phi$ are defined as in Fig.~\ref{fig:fcc}(a).
} \label{fig:spinwavez}
\end{figure}

Figure~\ref{fig:spinwavez} shows the quantum correction to the ground state energy for $J_3=1$ and $J_2=0.2$. $\theta$ and $\phi$ are defined as in Fig.~\ref{fig:fcc}(a) and Table~\ref{tab:fcc}. We find that single-$\bf q$ collinear antiferromagnetic (AFM) states [$\theta=0$ or $(\theta,\phi)=(\frac\pi2,0)$] are favored over multi-$\bf q$ states. Specifically, the staggered magnetization in each quantum ground state is oriented along a $\langle 100\rangle$ axis. This is a natural result as the fluctuations tend to favor collinear orders.  

\begin{figure}
\includegraphics[width=\linewidth]{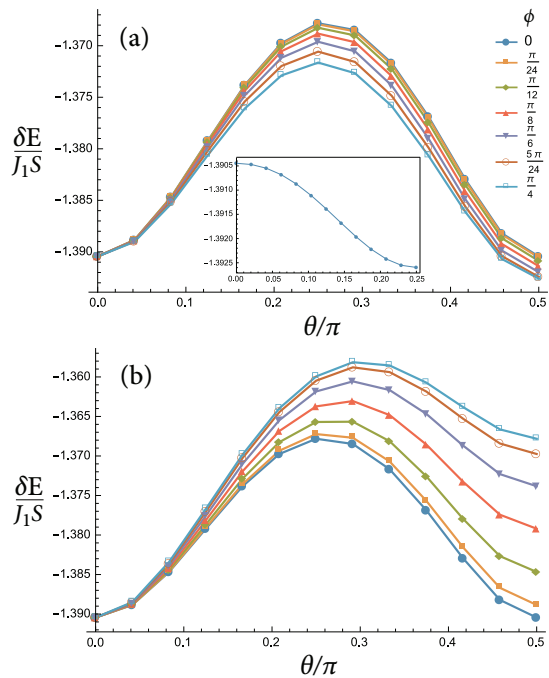}
\caption{
Quantum correction to the ground state energy for different ground states calculated by the spin wave analysis at $J_1=1$, $J_2=0.2$, and $J_3=-1$: (a) double-$\bf q$ and (b) triple-$\bf q$ magnetic ground states. $\theta$ and $\phi$ in (a) and (b) are defined as in Figs.~\ref{fig:fcc}(b) and \ref{fig:fcc}(c), respectively. The inset in (a) shows $\phi$ dependence of $\delta E/J_1S$ at $\theta=\pi/2$.
} \label{fig:spinwave}
\end{figure}

A similar behavior is also found for $J_3<0$. Figure~\ref{fig:spinwave}(a) shows the result for the double-$\bf q$ state [Fig.~\ref{fig:fcc}(b)]. The results indicate that single-$\bf q$ collinear AFM states ($\theta=0, \pi/2$) are favored over double-$\bf q$ orders. Furthermore, fluctuations lift the U(1) degeneracy of the single-$\bf q$ state at $\theta=\pi/2$. The inset in Fig.~\ref{fig:spinwave}(a) shows the quantum correction at $\theta=\pi/2$; the transverse axis is $\phi/\pi$. Here, $\phi=0$ corresponds to the collinear state with the spins pointing along the $\langle100\rangle$ direction, while they point along $\langle110\rangle$ for $\phi=\pi/4$. The minimum is located at $(\theta,\phi)=(\pi/2,\pi/4)$, indicating the single-$\bf q$ state with ${\bf q}=(0,0,2\pi)$ is favored by quantum fluctuations; the moments point along $(110)$ direction.

Estimation of the quantum correction for the triple-$\bf q$ ground states are shown in Fig.~\ref{fig:spinwave}(b). Similarly, the results indicate that, in the triple-$\bf q$ manifold, collinear states at $\theta=0$ and $(\theta,\phi)=(\pi/2,\pi/4)$ have the lowest ground state energy. This is the $\langle100\rangle$ collinear AFM state mentioned above. However, all of the states have higher energy than the $\langle110\rangle$ collinear phase. Hence, quantum fluctuations select $\langle110\rangle$ collinear order.

The favoring of the $\langle110\rangle$ direction for $J_3<0$ is a highly quantum effect.   It is very different from conventional magnetic anisotropy induced by crystal field effects.  In the latter case, a Landau theory analysis in powers of the (staggered) magnetization generally applies, and the leading effect occurs at fourth order.  These terms favor either $\langle100\rangle$ or $\langle111\rangle$, depending on the sign, but {\em not} $\langle110\rangle$.  

The above discussion pertains to the direction of the staggered magnetization, but we may also ask about the uniform susceptibility $\chi({\bf n})$ (${\bf n}={\bf h}/|h|$) with the four-sublattice magnetic structure. Expanding the energy in small deviations from the ground state, we find that $\chi({\bf n})$ for a given configuration is given by $\chi({\bf n})=\sum_{i,\alpha} m_{i\alpha}^2/\varepsilon_{i\alpha}$, where  $\varepsilon_{i\alpha,j\beta}=\partial^2 \varepsilon/\partial\mu_{i\alpha}\partial\mu_{j\beta}$ and $m_{i\alpha}=\partial m/\partial\mu_{i\alpha}$ with $\varepsilon$ and $m$ being ground state energy and magnetization along $\bf n$, respectively. Here, $\mu_{i\alpha}$ gives the small deviation of the $i$th moment along the orthogonal direction $\alpha=1,2$. Here, we defined the basis of $\mu_{i\alpha}$ so that $\varepsilon_{i\alpha,j\beta}$ become orthogonal, i.e., $\varepsilon_{i\alpha,j\beta}=\varepsilon_{i\alpha}\delta_{i,j}\delta_{\alpha,\beta}$. Using this formula, $\chi({\bf n})$ for the collinear states are given by $\chi({\bf n})=\sin^2\theta_{\bf n}/(4+J_3)$, as expected for collinear magnetic orders. Here, $\theta_{\bf n}$ is the angle between the collinear moments and $\bf n$. This indicates that, for $J_3>0$, the maximum of susceptibility will not be along $\langle111\rangle$.

To recapitulate the order by disorder analysis, we found that, within linear spin wave theory, quantum fluctuations select collinear ground states of the frustrated fcc antiferromagnet, and in these states the local moments are oriented along the $\langle 100\rangle$ or $\langle 110\rangle$ axis, depending upon parameters. The magnetic susceptibility, on the other hand, is maximal in $\{100\}$ planes for $J_3>0$. As we will see, this behavior contrasts that of the magnetic response in the paramagnetic phase, which is favors the $\langle111\rangle$ directions.

\underline{Paramagnetic response:} We next study the magnetic anisotropy in the paramagnetic phase. For this purpose, we evaluate the free energy per spin $f$ in an external magnetic field. We carry out a high temperature expansion of energy $f$ in powers of $\beta J_i$, where $\beta$ is the inverse temperature and $i=1,2,3$. The expansion of the partition function gives
\begin{eqnarray}
Z = Z_h \left< \exp\left\{ -\beta H_0 -\frac{\beta^2}2 [H_0,H_h] + {\cal O}[(\beta J_i)^3] \right\}\right>_h \label{eq:Zh}
\end{eqnarray}
where $Z_h=\text{Tr}\exp(-\beta H_h)$ and $\langle \hat{O} \rangle_h =Z_h^{-1}\text{Tr}\,\hat{O}\exp(-\beta H_h)$. As the contribution to $Z$ from the commutator in Eq.~\ref{eq:Zh} is in the order of ${\cal O}[(\beta J_i)^3]$, up to the second order in the expansion, we can ignore this term. Hence, $Z$ is approximated by
\begin{eqnarray}
Z = Z_h \left\{ 1 - \beta\langle H_0\rangle_h + \frac{\beta^2}2\langle H_0^2\rangle_h + {\cal O}[(\beta J_i)^3] \right\}. \label{eq:Zh2}
\end{eqnarray}
The second order correction to the free energy is given by
\begin{eqnarray}
\beta f^{(2)}&& (\beta {\bf h}) = (\beta J_1)^2 f^{(2)}_{0}(J_2/J_1,\beta h)\nonumber\\
&&+ (\beta J_3)^2 \left\{ f^{(2)}_{3} (\beta h) + g^{(2)}_{3} (\beta h) \sum_\alpha \hat{h}_\alpha^4 \right\},
\end{eqnarray}
where, $h=|{\bf h}|$ is the strength of external field, and $\hat{h}_\alpha = h_\alpha/h$ is the direction of external field. 
The contribution to the anisotropic term $g^{(2)}_{3} (\beta h)$ of the free energy comes only from $\langle[\beta J_3 \sum_{\langle i,j\rangle} ({\bf S}_i\cdot{\boldsymbol \delta}_{ij})({\bf S}_j\cdot{\boldsymbol \delta}_{ij})]^2\rangle$. The explicit form of $g^{(2)}_{3} (\beta h)$ is given by
\begin{eqnarray}
g^{(2)}_{3} (\beta h) &=& \frac{1}{16} \left\{S(S+1)-3\langle S^z{}^2\rangle_h+2\langle S^z\rangle_h{}^2 \right\}^2.\label{eq:g3}
\end{eqnarray}
Here, $\langle S^z{}\rangle_h$ and $\langle S^z{}^2\rangle_h$ are the thermal averages of $S^z$ and $S^z{}^2$, respectively, which are functions of the external field $\beta h$ and the spin size $S$. Because $g^{(2)}_{3} (\beta h)\ge0$, the result indicates that free energy is minimized when the external field is applied along $\langle111\rangle$.  Hence, we obtain $\langle111\rangle$ easy axis magnetic anisotropy regardless of the sign of $J_3$. We note that this anisotropy is present even for $S=1/2$. In this case, Eq.~(\ref{eq:g3}) becomes
\begin{eqnarray}
\left. g^{(2)}_{3} (\beta h)\right|_{S=1/2} = \frac1{64} \tanh^4 (\beta h/2).
\end{eqnarray}

Hence, for the model in Eq.~(\ref{eq:H}), spins favor the $\langle111\rangle$ axis in high temperature. On the other hand, for $J_3>0$ ($J_3<0$), the ground state is a collinear antiferromagnet with spins oriented along the $\langle100\rangle$ ($\langle110\rangle$) axes.  The distinct difference of the preferred axes at low and high temperatures constrasts with the common behavior of magnetic anisotropy, which is consistent at all temperatures. This can be understood as a consequence of the nature of fluctuations that dominate in different temperature regions. When $T\ll J_1$, long wave-length fluctuations dominate, lifting the accidental degeneracy of the ground states. On the other hand, in the high temperature limit, the spins are strongly disordered, and the spin correlation is limited to short range. Hence, in this regime, short range correlations between nearby spins is most important. Due to the difference in the nature of dominant fluctuations, the consequent phenomena can be different in the two regimes. 

This understanding also indicates that the information on direction-dependent spin interactions is reflected in the high temperature paramagnetic phase in an unusual way. Hence, this anisotropy-switching phenomena is potentially useful as a probe to experimentally narrow down the effective model for heavy element magnets, where various unconventional interactions can appear due to strong spin-orbit interactions.

Experimentally, a promising candidate to observe anisotropy switching is the family of double-perovskite compounds with heavy element ions.~\cite{Stitzer2002,Wiebe2002,Wiebe2003,Yamaura2006,Cussen2006,Erickson2007,deVries2010,Aharen2010} In these materials, strong spin-orbit coupling can induce strongly anisotropic interactions. Another favorable aspect is the suppression of magnetic ordering by geometrical frustration. As the magnetic anisotropy in the paramagnetic phase arises only at higher order in the inverse temperature, a low transition temperature should be favorable to enhance these higher order effects.

HI was supported by JSPS Postdoctoral Fellowships for Research Abroad. LB was supported by the NSF through grant NSF-DMR-12-06809.

\end{document}